\input{aipcheck}

\documentclass[
    ,final            
  ]
  {aipproc}

\layoutstyle{6x9}

\begin{document}

\title{Pulsar Science with the Green Bank 43m Telescope}


\author{M. B. Mickaliger}{
  address={Department of Physics, West Virginia University, Morgantown, WV 26506}
}

\author{M. A. McLaughlin}{
  address={Department of Physics, West Virginia University, Morgantown, WV 26506}
  ,altaddress={National Radio Astronomy Observatory, Green Bank, WV 24944}
}

\author{D. R. Lorimer}{
  address={Department of Physics, West Virginia University, Morgantown, WV 26506}
  ,altaddress={National Radio Astronomy Observatory, Green Bank, WV 24944}
}

\author{G. Langston}{
  address={National Radio Astronomy Observatory, Green Bank, WV 24944}
}

\author{A. V. Bilous}{
  address={Department of Astronomy, University of Virginia, Charlottesville, VA 22904}
}

\author{V. I. Kondratiev}{
  address={Netherlands Institue for Radio Astronomy (ASTRON), Postbus 2, 7990 AA Dwingeloo, The Netherlands}
  ,altaddress={Astro Space Center of the Lebedev Physical Institute, Profsoyuznaya str. 84/32, Moscow 117997, Russia}
}

\author{S. M. Ransom}{
  address={National Radio Astronomy Observatory, Charlottesville, VA 22903}
}

\author{F. Crawford}{
  address={Department of Physics and Astronomy, Franklin and Marshall College, Lancaster, PA 17604}
}

\begin{abstract}
The 43m telescope at the NRAO site in Green Bank, WV has recently been outfitted with a clone of the Green Bank Ultimate Pulsar Processing Instrument (GUPPI \cite{Ransom:2009}) backend, making it very useful for a number of pulsar related studies in frequency ranges 800-1600 MHz and 220-440 MHz. Some of the recent science being done with it include: monitoring of the Crab pulsar, a blind search for transient sources, pulsar searches of targets of opportunity, and an all-sky mapping project. For the Crab monitoring project, regular observations are searched for giant pulses (GPs), which are then correlated with $\gamma$-ray photons from the \emph{Fermi} spacecraft. Data from the all-sky mapping project are first run through a pipeline that does a blind transient search, looking for single pulses over a DM range of 0-500 pc~cm$^{-3}$. These projects are made possible by MIT Lincoln Labs.
\end{abstract}

\maketitle


\section{Transient Detection Pipeline}

The ongoing observations of the Crab pulsar are being reduced using a new real-time pipeline. The data are taken with WUPPI and processed on Tofu, which has 16 processor cores, allowing 16 processes to be run in parallel. This pipeline is comprised of a set of scripts using freely available analysis tools (http://sigproc.sourceforge.net) that can reduce a file in the same amount of time it is taken in. The pipeline outputs profiles for each GP, observation time vs pulse phase plots, a folded profile, and an average GP profile, made by summing all of the individual GP profiles.

\section{\emph{Fermi} Data}

\emph{Fermi} data are downloaded from the online archive (\url{http://fermi.gsfc.nasa.gov/cgi-bin/ssc/LAT/LATDataQuery.cgi}) for times during which radio observations occured. This amounts to about 5400 $\gamma$-ray photons over all of the days processed so far. The $\gamma$-ray photons for each day ($\sim$180) are folded to make sure the Crab is visible, then they are correlated with the GP TOAs. Correlations are searched for with time lags of 1 pulsar spin period, 1 second, and every 10 seconds up to 60 seconds.

\section{$\gamma$-ray / Radio Correlation}

\begin{figure}[hb]
  \begin{tabular}{cc}
    \frame{\includegraphics[height=.5\textheight, width=0.248\textheight, angle=-90]{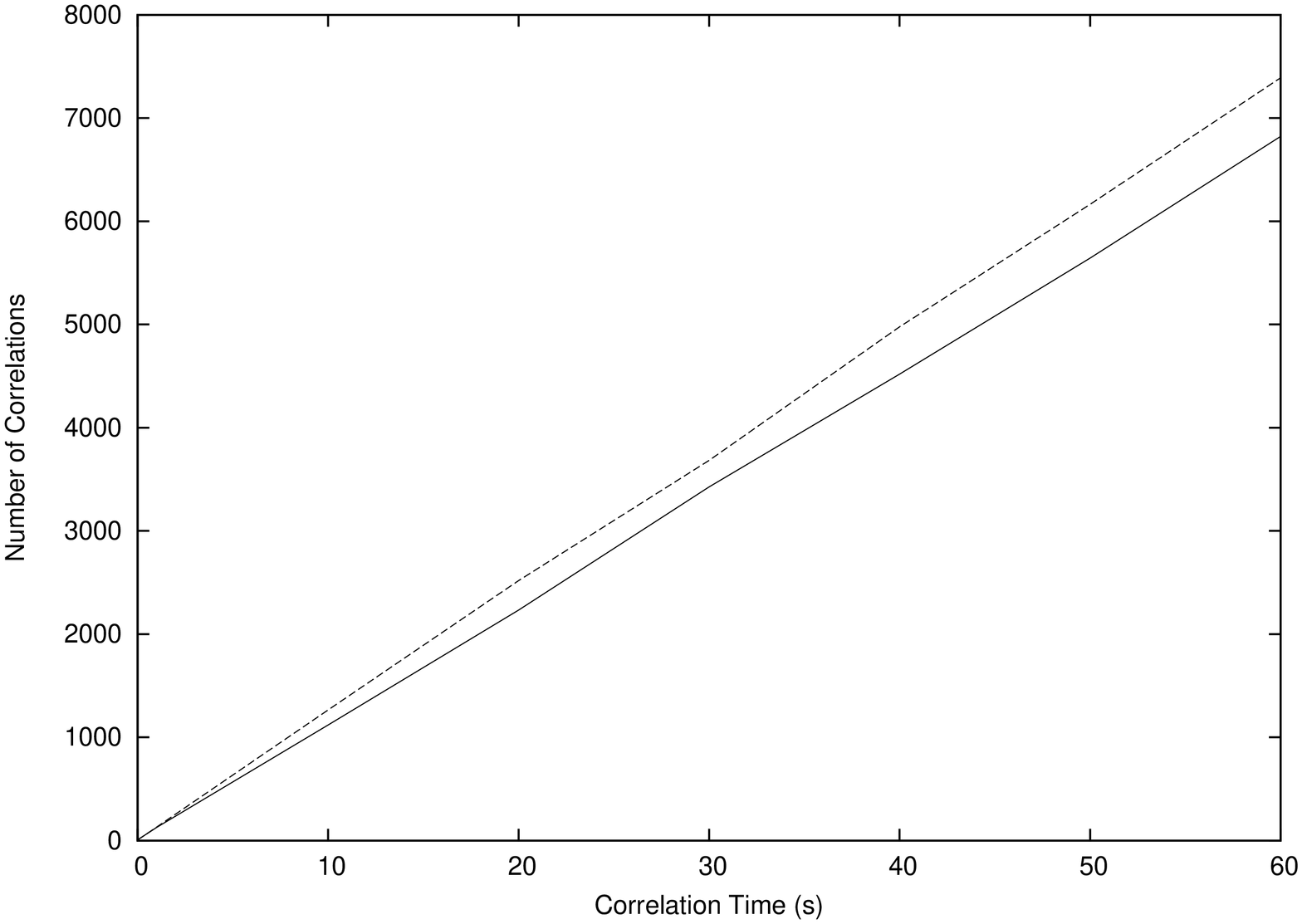}} \\
    \frame{\includegraphics[height=.5\textheight, width=0.248\textheight, angle=-90]{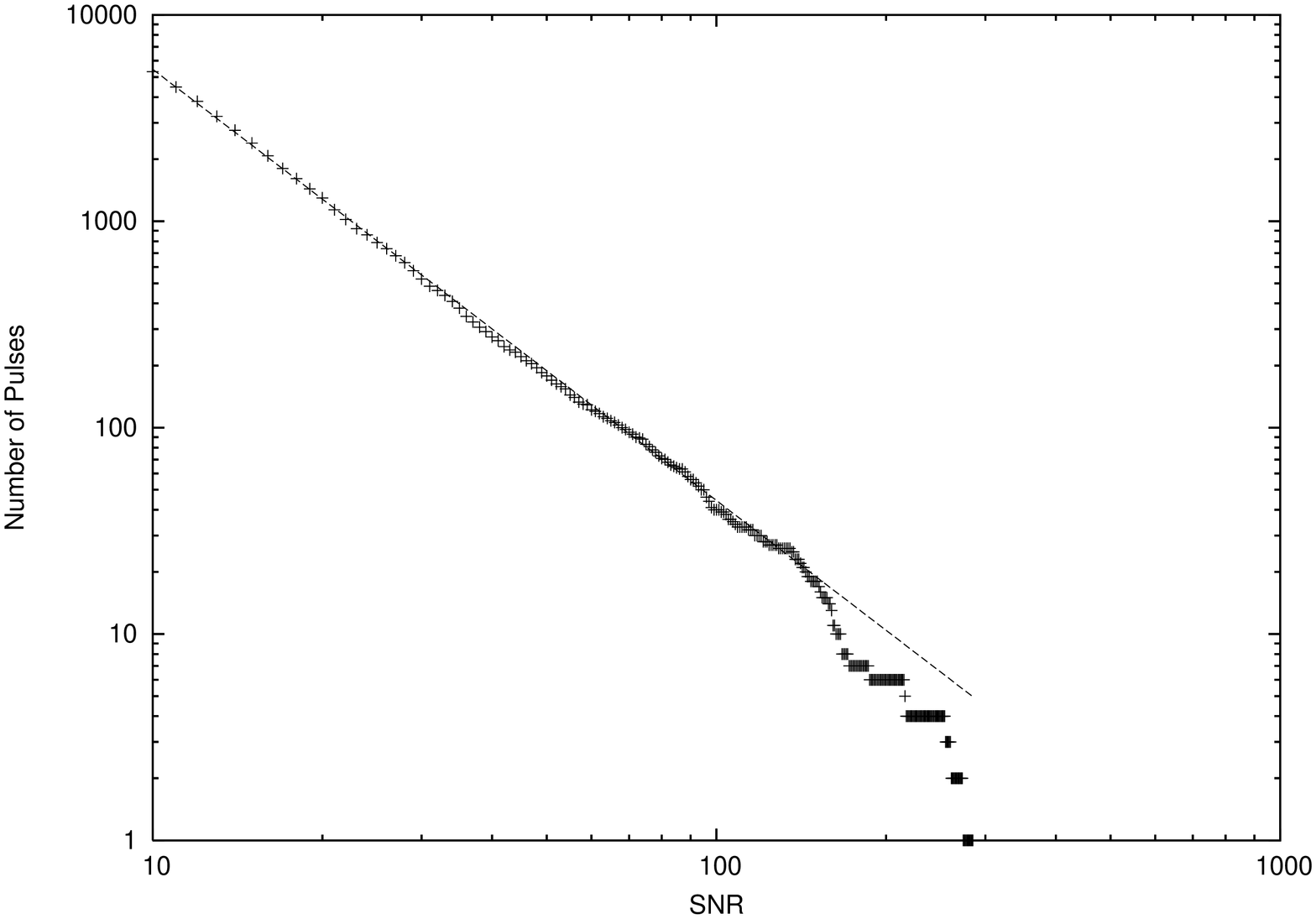}}
  \end{tabular}
  \caption{{\bf Top:} Correlation between radio GPs and $\gamma$-ray photons (solid line). Also plotted is a correlation between $\gamma$-ray photons and randomized radio TOAs (dashed line). {\bf Bottom:} Amplitude distribution and power-law fit, with $\alpha$=2.09.}
\label{fig:fermi}
\end{figure}

As shown in Figure \ref{fig:fermi} (top), randomized TOAs correlate just as well as real data do with \emph{Fermi} photons. This lack of significant correlation means that increased coherence or changes in beaming seem to be the cause of GPs, as opposed to an increase in pair production in the magnetosphere.

In total, $\sim$23000 GPs have been analyzed to date. A cumulative amplitude distribution (Figure \ref{fig:fermi}, bottom) for a subset of those pulses was made. This can be fit by a power-law with slope = -2.09, in agreement with other published values (e.g. \cite{Cordes:2004}).


\bibliographystyle{aipproc}   

\end{document}